\documentclass{amsart}
\newtheorem{lemma}{Lemma}[section]

\newtheorem{corollary}[lemma]{Corollary}
\newtheorem{theorem}[lemma]{Theorem}
\theoremstyle{definition}

\theoremstyle{definition}
\newtheorem{example}[lemma]{Example}
\newtheorem{remark}[lemma]{Remark}
\theoremstyle{definition}

\thanks{This work was supported by Science and Technology Assistance
Agency under the contract No. APVT-51-032002, grant VEGA 2/3163/23
and Center of excellence SAS, CEPI I/2/2005}
\title {How sharp are PV measures?}
\author{Jen\v cov\'a, A., Pulmannov\'a, S.}
\begin{document}
\begin{abstract}
Properties of sharp observables (normalized PV measures) in relation
to smearing by a Markov kernel  are studied. It is shown that for a
sharp observable $P$ defined on a standard Borel space, and an
arbitrary observable $M$, the following properties are equivalent:
(a) the range of  $P$ is contained in the range of $M$; (b) $P$ is a
function of $M$; (c) $P$ is a smearing of $M$.
\end{abstract}
\address{Mathematical Institute, Slovak Academy of Sciences, \v
Stef\'anikova 49, 814 73 Bratislava, Slovakia }
\email{pulmann@mat.savba.sk, jenca@mat.savba.sk}  \maketitle
\markboth{Jen\v cov\'a, A., Pulmannov\'a, S.} {How sharp are PV
measures?}
\date{}
\maketitle

\section{Introduction}

Normalized POV (positive operator valued) measures are used to
describe generalized observables in quantum mechanics (\cite{L, Hol,
BGL}). Their introduction is justified by the analysis of some ideal
experiments which shows that there are quantum events that cannot be
described by projections \cite{BGL}. POV measures are also used to
generalize Mackey's imprimitivity  theorem \cite{AE, T} and to study
the problem of the joint measurements of incompatible observables
\cite{MM, Uf, CHT, W}.

Generalized yes-no experiments are in one-to one correspondence with
self-adjoint operators lying between $0$ and $I$ (with respect to
the usual ordering of self-adjoint operators). These operators are
called quantum effects. Let ${\mathcal E}(H)$ denote the set of all
quantum effects on a Hilbert space $H$, i.e., ${\mathcal E}(H):=\{
T: 0\leq T\leq I\}$, where $T$ is a self-adjoint operator.
Projection operators are contained in ${\mathcal E}(H)$, and they
are distinguished among the effects by the equality $P\wedge
(I-P)=0$, which can be interpreted as the property that events $P$
and non-$P$ cannot simultaneously occur. Projection operators are
called {\it sharp} effects, while the other effects are {\it
unsharp}. Correspondingly, PV (projection valued) observables are
called {\it sharp} observables \cite{AD}.

Recall that the states on ${\mathcal E}(H)$ (i.e. the physical
states of the corresponding physical system) coincide with the set
of all density operators on $H$.  There exists a one-to-one
correspondence between POV measures (defined on a measurable space
$(X,{\mathcal B})$) and affine maps from the set of states into the
set of probability measures on $(X,{\mathcal B})$, which is based on
the interpretation of the number $Tr[SF(\Delta)]$ as the probability
that the outcome of a measurement of the observable (POV measure)
$F$ is in $\Delta \in {\mathcal B}$ if the physical system is in the
state $S$ \cite{Hol}. This one-to-one correspondence allows one to
apply some results of the classical mathematical statistics to
quantum experiments \cite{Hol,JPV}. In particular, given a
probability measure $\mu$ and a suitable Markov kernel $\lambda$, we
can form another probability measure, $\lambda\circ \mu$, so-called
randomization of $\mu$ by $\lambda$ \cite{St}. This has been applied
to quantum observables: to a given observable and a suitable Markov
kernel, a new observable can be created, which is called a smearing,
or a fuzzy version, of the given observable \cite{Hol, HLY, H1,
JPV}. For example, it is well known that an unsharp observable is a
smearing of a sharp observable iff its range is commutative
\cite{Hol1, B, CN, JPV}. A partial ordering can be introduced on the
set of observables by defining $E\preceq F$ if the observable $F$ is
a smearing of the observable $E$ \cite{BDKPW, H, JPV}. Minimal
points in this ordering are called clean observables \cite{BDKPW}.

In the present paper, we study properties of sharp observables in
relation to smearing. In our considerations, we often replace a
Markov kernel by a weak Markov kernel to simplify the proofs, and
then apply well known results about the equivalence of  the weak
Markov kernel with its regular version, which is a Markov kernel. We
show that a sharp observable $P$, defined on a standard Borel space,
can be considered as a smearing of another, in general unsharp
observable $M$, iff the corresponding Markov kernel is of a special
type, which makes the sharp observable $P$ a function of the unsharp
observable $M$. We also show that this holds not only for sharp
observables, but for all observables which are extremal with respect
to the convex structure of observables. Consequently, a sharp
observable is clean iff its range generates a maximal abelian von
Neumann subalgebra of the bounded operators on $H$. We also show
that for a sharp observable $P$ defined on a standard Borel space,
and an arbitrary observable $M$, the following properties are
equivalent: (a) the range of  $P$ is contained in the range of $M$;
(b) $P$ is a function of $M$; (c) $P$ is a smearing of $M$. We note
that the equivalence of (a) and (b) has been proved in \cite{DLPY},
where the Naimark theorem was used. In this paper, we give a
different proof.

\section{Smearing of observables}

Let $H$ be a (complex, separable) Hilbert space. Let ${\mathcal
E}(H)$ be the set of effects on $H$ and let ${\mathcal S}$ be the set of states on
${\mathcal E}(H)$. We recall the following
property of the order on $\mathcal E(H)$, inherited from the usual
order on self-adjoint operators:
\begin{equation}
a\le b \quad \mbox{if and only if}\quad ab=ba=a
\end{equation}
whenever $a,b\in \mathcal E(H)$ and $a$ or $b$ is a projection.

Let $(X, {\mathcal A})$ and $(Y, {\mathcal B})$ be measurable spaces
and let $E:(X,{\mathcal A}) \to {\mathcal E}(H)$  be  a POV measure.
Assume further that there is a map  $\lambda :X\times {\mathcal B}
\to [0,1]$ such that
\begin{enumerate}
\item[(i)] $\lambda(.,B)$  is ${\mathcal
A}$-measurable for all $B\in {\mathcal B}$,
\item[(ii)]  $\lambda(x,.)$ is a
probability measure on ${\mathcal B}$ for all $x\in X$.
\end{enumerate}
 That is, $\lambda$ is a Markov kernel. Then
\begin{equation}\label{eq:1}
\lambda\circ E(B):=\int_X \lambda(x,B)E(dx),\qquad B\in \mathcal B
\end{equation}
defines a POV - measure $(Y,\mathcal B)\to \mathcal E(H)$, called the
 {\it smearing} of $E$ with respect to $\lambda$.

The notion of a Markov kernel can be weakened as follows.  Let
${\mathcal P}\subseteq M_1^+(X, {\mathcal A})$, where $M_1^+(X,
{\mathcal A})$ denotes the set of probability measures on
$(X,{\mathcal A})$, and let $\nu: X \times {\mathcal B} \to {\mathbb
R}$. We will say that $\nu$ is a {\it weak Markov kernel with
respect to ${\mathcal P}$} if
\begin{enumerate}
\item[(i)] $x \mapsto \nu(x, B)$ is ${\mathcal
A}$-measurable for all $B\in {\mathcal B}$;
\item[(ii)] for every $B\in {\mathcal B}$, $0\leq \nu(x,
B)\leq 1,\,  {\mathcal P}$-a.e.;
\item[(iii)] $\nu(x, Y)=1,\,  {\mathcal P}$-a.e. and
$\nu(x,\emptyset)=0,\, {\mathcal P}$-a.e. .
\item[(iv)] if $\{ B_n\}$ is a sequence in ${\mathcal B}$ such
that $B_n\cap B_m=\emptyset$ for $m\neq n$, then
$$
\nu(x, \bigcup_n B_n)=\sum_n \nu(x, B_n),\, {\mathcal
P}-a.e. .
$$
\end{enumerate}

Let $E$ be   as above and put ${\mathcal P}= \{ m\circ E: m\in {\mathcal S}\}$. If
$\nu:X\times {\mathcal B}\to {\mathbb R}$ is a weak Markov kernel
with respect to $\mathcal P$,  then we will say that $\nu$ is a weak Markov kernel
with respect to $E$ and
$$
\nu\circ E(B):=\int_X\nu(x,B)E(dx), \qquad B\in \mathcal B
$$
defines a POV - measure, which will be called a  smearing of $E$ with respect to $\nu$.

\begin{remark}\label{re:reg} We note that a weak Markov kernel
$\nu:X\times {\mathcal B}\to[0,1]$ (with respect to one probability
measure $P$) is called a {\it random measure} in the literature. If
${\mathcal B}$ is the Borel $\sigma$-algebra of subsets of a
complete separable metric space $Y$, then there exists a regular
version $\nu^*$ of $\nu$, such that $\nu^*$ is a Markov kernel, and
\begin{equation}\label{eq:reg} \forall B\in {\mathcal B},
\nu(x,B)=\nu^*(x,B),\,\, a.e. P
\end{equation}
(see, e.g. \cite[VI.1. 21.]{St}).

Notice further that since on a separable Hilbert space, there exists
a faithful state $m_0\in {\mathcal S}$, and $m\circ E$ is dominated
by $m_0\circ E$ for all $m\in {\mathcal S}$, then $\nu$ is a weak
Markov kernel with respect to $\{ m\circ E: m\in {\mathcal S}\}$ iff
$\nu$ is a weak Markov kernel with respect to $m_0\circ E$.

Moreover, it has been proved in \cite{JPV} that if an observable $F:
(Y,{\mathcal B})$ is a smearing of an observable $E$ with respect to
a weak Markov kernel $\nu$, and $(Y,{\mathcal B})$ is a standard
Borel space, then there is a Markov kernel $\nu^*$ such that $F$ is
a smearing of $E$ with respect to $\nu^*$.

\end{remark}

\section{PV - measures and smearings}

For an observable $E$, let $\mathcal R(E)$ denote the range of $E$.
The following Theorem is well known, see \cite{Hol1, B, CN, JPV}.
For completeness, we include (a sketch of) the proof, as it was
given in \cite{JPV}.

\begin{theorem}\label{thm:commut} Let $M:(Y,\mathcal B)\to \mathcal E(H)$
be a POV - measure. Then $M$ is a smearing
of some  PV - measure $P$ with respect to a weak Markov kernel if and only if
$\mathcal R(M)$ is commutative.
\end{theorem}

\begin{proof}
Since $\mathcal R(M)$ is commutative, there is a self-adjoint
operator $T$ on $H$, such that $\mathcal R(M)\subset \{T\}''$. It
follows that for each $B\in \mathcal B$, there is a Borel function
$f_B$, such that
$$
M(B)=f_B(T)=\int_{\mathbb R} f_B(x)P(dx),
$$
 where $P$ is the spectral measure of $T$. It is not difficult to show that $\nu(x,B)=f_B(x)$ defines a
 weak Markov kernel $X\times \mathcal B\to \mathbb R$ with respect to $P$.

 The converse statement is obvious.

\end{proof}

The main purpose of this paper is to study the opposite situation, namely when a PV - measure $P$
is a smearing of some observable $M$.

\begin{theorem}\label{thm:3to1} Let $M:(X,\mathcal A)\to \mathcal E(H)$ be a POV - measure and let $P: (Y,\mathcal B)\to \mathcal E(H)$
 be a PV - measure. Let $\nu: X\times B\to \mathbb R$ be a weak Markov kernel  with respect to $M$ and suppose that
 $P=\nu\circ M$. Then $\mathcal R(P)\subset \mathcal R(M)$.
\end{theorem}

\begin{proof} Let $B\in \mathcal B$, then
 $P(B)=\int \nu(x,B)M(dx)$. Put $\pi(B):=\{ x:\nu(x,B)=1\}$.
Let $m$ be a state on $\mathcal E(H)$, with the support $supp(m)=P(B)$. Then
$$
1=m(P(B))=\int \nu(x,B)m(M(dx))
$$
hence $m\circ M(\pi(B)^c)=0$. Since $P(B)$ is the support of $m$, we
have $P(B)M(\pi(B)^c)P(B)=0$. By positivity of $M$ this entails
$$P(B)M(\pi(B)^c)=M(\pi(B)^c)P(B)=0$$
 Therefore $P(B)M(\pi(B))=M(\pi(B))P(B)=P(B)$, hence $P(B)\leq M(\pi(B))$. Similarly, $P(B^c)\leq M(\pi(B^c))$. But
$$
I=P(B)+P(B^c)=M(\pi(B))+M(\pi(B)^c)
$$
yields $M(\pi(B)^c)\leq P(B^c)$,
and since, by definition, $\pi(B^c)\subseteq \pi(B)^c$ modulo $M$, we get
$$
P(B^c)\leq M(\pi(B^c))\leq M(\pi(B)^c)\leq P(B^c).
$$
We conclude that $P(B^c)=M(\pi(B^c))=M(\pi(B)^c)$, and therefore
$P(B)=M(\pi(B))\in {\mathcal R}(M)$.
\end{proof}

As an example, we will consider in details the case of a finite
dimensional Hilbert space.

\begin{example}\label{ex:fin}Let $H$ be finite dimensional. Let
$Y$ be a finite set  and let $P:Y\to {\mathcal E}(H)$ be a PV
measure.  Assume that $P=\nu\circ M$ with  a weak Markov kernel
$\nu$ and a POV - measure $M: (X,{\mathcal A})\to {\mathcal E}(H)$.
Since $Y$ is finite, there is a set $C\subset Y$, such that the
restriction of $\nu$ to $C^c$ is a Markov kernel and $M(C)=0$.

For $y\in Y$,  put $\pi(y):=\{  x\in C^c: \nu(x,y) = 1\}$. As in the above Theorem,
$P(y)=M(\pi(y))$. Moreover, since  $\sum_{y\in Y}\nu(x,y)=1$ for $x\in C^c$, we obtain that
$x\in \pi(y)$ implies that  $\nu(x,y')=0$ and therefore $x\in \pi(y')^c$, for $y'\neq y$.
This shows  that $\{ \pi(y): y\in Y, C\}$ is a partition of $Y$.

Moreover, we can define
a Markov kernel $\nu^*:X\times \mathcal B\to [0,1]$ by
$$
\nu^*(x,y)=\left\{ \begin{array}{r@{\quad:\quad}l}  \chi_{\pi(y)}(x)& x\in C^c\\
        \mu(y)& x\in C
        \end{array}\right.
$$
where $\mu$ is any probability measure on $Y$.
Then $\nu(x,y)=\nu^*(x,y)$ for $x\in C^c$ and we have $P= \nu^*\circ M$.
The observable $M$ has the following form: if $H_y:=P(y)H$, then $H=\oplus_y H_y$
and
$$
M(A)=\oplus_y P(y)M(A)=\oplus_y M(A\cap\pi(y))
$$

\end{example}

In the above example, note that the weak Markov kernel must satisfy
$\nu(x,y)\in \{0,1\}$ for $y\in Y$ and all $x$ in $C^c$. More
generally, if $\nu: X\times \mathcal B\to \mathbb R$ is a weak
Markov kernel with respect to a POV measure $M$, we will say that
$\nu$ has values in $\{0,1\}$ if for each $B\in\mathcal B$,
$\nu(x,B)\in \{0,1\}$, a.e. - $\{m\circ M,\ :\ m\in \mathcal S\}$.

Let $(X,{\mathcal A})$ be a measurable space, and let
$E_i:(X,{\mathcal A})\to {\mathcal E}(H), \ i=1,2$ be POV measures.
For every $\alpha \in [0,1]$, $A\mapsto E(A)=\alpha
E_1(A)+(1-\alpha)E_2(A)$, $A\in {\mathcal A}$, defines a POV
measure. Hence the set of all observables associated with
$(X,{\mathcal A})$ bears a convex structure. Since projections are
extremal points in the convex set ${\mathcal E}(H)$, sharp
observables are  extremal  in the set of all observables associated
with a given measurable space. In general, however, there exist
extremal points which  are unsharp, \cite{Hol2}.

\begin{theorem}\label{th:extrem} Let $E$ be a  POV measure which is an extreme
point in the convex set of all  POV measures defined on a
measurable  space $(X,{\mathcal A})$. Then if $E$ is a smearing of a POV
measure $M$ with respect to a weak Markov kernel $\nu$, then $\nu$ has
values in $\{0,1\}$. Moreover, if $(X,\mathcal A)$ is a standard Borel space,
then $E$ is a function of $M$.
\end{theorem}
\begin{proof} Let  $M:(Y,{\mathcal B})\to {\mathcal E}(H)$ be a POV measure
and let  $\nu :Y\times
{\mathcal A}\to\mathbb R$ be a weak Markov kernel with respect to $M$.
Suppose that $E = \nu\circ M$.

Fix $B_1\in {\mathcal A}$.  Define
$$
\nu^{\pm}(y,B):=\nu(y,B)\pm [\nu(y,B_1)\nu(y,B\cap
B_1^c)-\nu(y,B_1^c)\nu(y, B\cap B_1)].
$$
Then $\nu^{\pm}(y,B)$ is a weak Markov kernel with respect to $M$,
\cite{Hol}. Moreover,
$$
\nu(y,B)=\frac{1}{2}\nu^+(y,B)+\frac{1}{2}\nu^-(y,B).
$$
This implies that $E(B)=1/2E^+(B) + 1/2 E^-(B)$, where
$E^{\pm}(B)=\int_Y\nu^{\pm}(y,B)M(dy)$. Since $E$ is extremal,
we must have $E^+=E^-=E$, which implies that
$$
\int [\nu(y,B_1)\nu(y,B\cap B_1^c)-\nu(y,B_1^c)\nu(y,B\cap B_1)]M(dy)=0.
$$
Then
\begin{eqnarray*}
E(B)&=& \int \nu(y,B)M(dy)=\int(\nu(y,B\cap B_1)+\nu(y,B\cap B_1^c))M(dy)\\
&=& \int [\nu(y,B_1)\nu(y,B\cap B_1)+\nu(y, B_1^c)\nu(y,B\cap B_1)
+ \nu(y,B\cap B_1^c)]M(dy)\\
&=& \int (\nu(y,B_1)\nu(y,B\cap B_1)+\nu(y,B_1)\nu(y,B\cap B_1^c)
+\nu(y,B\cap B_1^c))M(dy)\\
&=& \int [ \nu(y,B_1)\nu(y,B)+\nu(y,B\cap B_1^c)]M(dy).
\end{eqnarray*}
In particular,
$$
E(B_1)=\int \nu(y,B_1)^2M(dy)=\int \nu(y,B_1)M(dy)
$$
and since $\nu(y,B_1)\geq \nu(y,B_1)^2$ a.e. $M$, we get
$\nu(y,B_1)=\nu(y,B_1)^2$ a.e. $M$, that is,
$\nu(y,B_1)\in \{ 0,1\}$ a.e. $M$. Since $B_1$ was arbitrary,
this holds for all $B\in {\mathcal A}$.

Suppose next that $(X,\mathcal A)$ is a standard Borel space.
Then there exists a Markov kernel $\lambda:Y\times \mathcal A\to [0,1]$,
such that $P=\lambda \circ M$.
Put $\pi(A):=\{ y: \lambda(y,A)=1\}$.

First, we will show that $E(A)=
M(\pi(A))$. For every $A$,  $\pi(A)\cap \pi(A^c)=\emptyset$, because
$\lambda(y,.)$ is a probability measure. Therefore, there is a partition
$Y=\pi(A)\cup \pi(A^c)\cup C_A$ and, by the first part of the proof,
$M(C_A)=0$. Then
$$
E(A)=\int_{\pi(A)} \lambda(y,A)M(dy) + \int_{\pi(A^c)} \lambda(y,A)M(dy)
+ \int_{C_A} \lambda(y,A)M(dy)
$$
The first integral is $M(\pi(A))$, the other two are $0$.

Next, we show  that $\pi$ is a $\sigma$-homomorphism of sets
modulo $M$.

(1) $\pi(A\cap B)=\{ y: \lambda(y, A\cap B)=1\}$. Since
$\lambda(y,.)$ is a probability measure, we have $\lambda(y,A)=
\lambda(y,B)=1$ if and only if $\lambda(y,A\cap B)=1$. This
entails $\pi(A\cap B)=\pi(A)\cap \pi(B)$.

(2)
Observe that $A\subset B$ implies $\pi(A)\subset \pi(B)$,
which follows from $\lambda(y,A)\leq \lambda(y,B)$.

(3) Let $A_n\in {\mathcal A}$, $n\in {\mathbb N}$, $A_n\cap
A_m=\emptyset$ if $n\neq m$. Put $A=\cup A_n$. Then $A_n\subset A \,
\Rightarrow \, \pi(A_n)\subset \pi(A)$, hence $\cup \pi(A_n)\subset
\pi(A)$. Let $C=\pi(A)\setminus \cup \pi(A_n)$. If $y\in C$, then
$\lambda(y,A)=1$ and $\lambda(y,A_n)\neq 1$ for all $n$, so that,
for all $n$, either $\lambda(y,A_n)=0$ or $y\in C_{A_n}$. Since
$\lambda(y,A)=\sum \lambda(y,A_n)=1$, there is an $n$ such that
$\lambda(y,A_n)\neq 0$, that is,  $y\in C_{A_n}$. Therefore
$C\subseteq \cup_n C_{A_n}$, so that $M(C)=0$. This concludes
the proof that $\pi$ is a set homomorphism modulo $M$.

Let $m$ be a faithful state on $E(H)$, $m\circ M=\mu$ is a
probability measure on ${\mathcal B}$. Put $I:=\{ B\in {\mathcal B}:
\mu(B)=0\}$, then $I$ is a $\sigma$-ideal, and ${\mathcal B}/I$ is a
Boolean $\sigma$-algebra. Let $p:B\mapsto [B]$ be the canonical
homomorphism. Put $\pi_1:{\mathcal A}{\overset \pi\to} {\mathcal
B}{\overset p\to}{\mathcal B}/I$. Then $\pi_1$ is a
$\sigma$-homomorphism. The triple $(Y,{\mathcal B},p)$, where
$p:{\mathcal B}\to {\mathcal B}/I$ is surjective, satisfies
conditions of  \cite[Lemma 4.1.8]{PtPu}, resp. \cite[Theorem
1.4]{Var}, and hence there is $f:Y\to X$, measurable and such that
$\pi_1(A)=p\circ f^{-1}(A)$, $A\in{\mathcal A}$.  By the definition
of $I$, and since $m$ is faithful, if $B_1\in [B]$, then
$M(B_1)=M(B)$. Hence $\pi_1(A)=[f^{-1}(A)]=[\pi(A)] \, \Rightarrow
\, E(A)=M(\pi(A))=M(f^{-1}(A))$.

\end{proof}

Next we will show the converse to Theorem \ref{thm:3to1}.

\begin{theorem}\label{th:(i)to(iii)} Let
$P: (Y,{\mathcal B})\to {\mathcal E}(H)$ be a PV measure and let $M:
(X,{\mathcal A})\to {\mathcal E}(H)$ be a POV measure. If ${\mathcal
R}(P)\subseteq {\mathcal R}(M)$, then
 there is a weak Markov kernel $\nu$ with respect to $M$, such that $P$ is
a smearing of $M$.
\end{theorem}
\begin{proof} The assumption  implies that there is a mapping $\pi :{\mathcal
B}\to {\mathcal A}$ such that $P(B)=M(\pi(B))$, $B\in {\mathcal B}$.
The latter equality entails that
$$
P(B)=\int_X \chi_{\pi(B)}M(dx), B\in {\mathcal B}.
$$

Put $\nu(x,B)=\chi_{\pi(B)}(x)$, $B\in {\mathcal B}$. We will
prove that $\nu :X\times {\mathcal B}\to [0,1]$ is a weak Markov
kernel with respect to $M$.

Clearly, $0\leq \nu(x,B)\leq 1$, $\nu(x,Y)=1$ a.e. $M$,
$\nu(x,\emptyset)=0$ a.e. $M$.

Let $\{ B_n\}_n$ be a sequence of elements in ${\mathcal B}$,
$B_m\cap B_n=\emptyset$, $m\neq n$, and denote $B:=\bigcup_n B_n$.
We have
$$
M(\pi(B))=P(B)=\sum_n P(B_n)=\sum_n M(\pi(B_n)).
$$

We will show that
$$
\sum_n M(\pi(B_n))=M(\bigcup_n \pi(B_n)).
$$

First, observe that
$B_1\cap B_2=\emptyset$ implies that
$M(\pi(B_1))M(\pi(B_2))=0$
and from $M(\pi(B_1)\cap \pi(B_2))\leq M(\pi(B_1)), M(\pi(B_2))$ we
derive that
$$
M(\pi(B_1)\cap \pi(B_2))=0.
$$

Consider the sequence $\{ C_n\}$, where
\begin{eqnarray*}
C_1&=&\pi(B_1),\\
\ldots\\
C_n&=&\pi(B_n)\setminus
\bigcup_{k=1}^{n-1}\pi(B_k)=\bigcap_{k=1}^{n-1}\pi(B_n)\cap \pi(B_k)^c\\
\ldots
\end{eqnarray*}

Then $C_n\cap C_m=\emptyset$, $n\neq m$, and $C_n\subseteq \pi(B_n)
\, \forall n$. In addition,
\begin{eqnarray*}
A_n:=\pi(B_n)\cap C_n^c&=&\pi(B_n)\cap(\bigcup_{k=1}^{n-1}\pi(B_n)^c\cup
\pi(B_k))\\
&=& \bigcup_{k=1}^{n-1}\pi(B_n)\cap \pi(B_k),
\end{eqnarray*}
which implies $\pi(B_n)=C_n\cup A_n$, $M(A_n)=0$, whence
$M(\bigcup_n \pi(B_n))=M(\bigcup_n C_n)=\sum_n M(C_n)=\sum_n
M(\pi(B_n))=M(\pi(B))$.

\smallskip
We will show that we can replace $C_n$ by a sequence $D_n$ such that
$D_n\subset \pi(B)$ $\forall n$. Clearly, $P(B_n)\leq P(B)$ for all
$n$. Since $M(\pi(B_n)\cap \pi(B)^c)\leq M(\pi(B_n))=P(B_n)\leq
P(B)$ and also $M(\pi(B_n)\cap \pi(B)^c)\leq M(\pi(B))^c=P(B)^c$, we
obtain
$$
M(\pi(B_n)\cap \pi(B)^c)=0.
$$

Since $C_n\subseteq \pi(B_n)$ we have  $M(C_n\cap \pi(B)^c)=0$. Put
$$
D_n:=C_n\cap \pi(B),\qquad F_n:=\pi(B_n)\setminus D_n= C_n\cap
\pi(B)^c\cup A_n $$ Then  $\pi(B_n)=D_n\cup F_n$, with $ M(F_n)=0$.
Moreover, $D_n\cap D_m=\emptyset, n\neq m$, and $D_n\subseteq
\pi(B), \forall n$. This entails $M(\bigcup D_n)=M(\bigcup
\pi(B_n))$, the left hand side equals $\sum_n M(D_n)=\sum_n
M(\pi(B_n))=M(\pi(B))$. Therefore
$$
\sum_{k=1}^{\infty}\chi_{\pi(B_k)}(y)=\sum_{k=1}^{\infty}\chi_{D_k}(y)+\sum_{k=1}^{\infty}\chi_{F_k}(y),
$$
where the second term on the right is equal to $0$ a.e. $M$.
Further,
$$
 \chi_{\pi(B)}(y)-\sum_{k=1}^n \chi_{\pi(B_k)}(y)=
 \chi_{\pi(B)}(y)-\sum_{k=1}^n\chi_{D_k}-\sum_{k=1}^n\chi_{F_k}(y)\geq
 0\  a.e. M.
 $$
From
$$
\int (\chi_{\pi(B)}(y)-\sum_{k=1}^n\chi_{\pi(B_n)}(y))M(dy) \to 0,
$$
we obtain, since the sub-integral function is bounded,

$$
\int \lim_n(\chi_{\pi(B)}(y)-\sum_{k=1}^n\chi_{\pi(B_k)}(y))M(dy)
=0,
$$
which implies $\sum_{n=1}^\infty\chi_{\pi(B_n)}(y)=\chi_{\pi(B)}(y)$ a.e.
$M$. This concludes the proof that $\nu(y,B)=\chi_{\pi(B)}(y)$
is a weak Markov kernel.

\end{proof}

Our results so far can be summarized as follows.

\begin{theorem}\label{th:equiv} Let $M:(X,{\mathcal A})\to E(H)$ be
a POV measure and   $P:(Y,{\mathcal B})\to {\mathcal E}(H)$ be a PV
measure.  The following conditions are equivalent:
\begin{enumerate}
\item[{\rm(i)}] ${\mathcal R}(P) \subset {\mathcal R}(M)$,
\item[{\rm(ii)}] there exists a weak Markov kernel $\nu$
with respect to $M$  with values in $\{0,1\}$, such that $P=\nu\circ M$,
\item[{\rm(iii)}] there exists a weak Markov kernel $\nu$
with respect to $M$, such that  $P=\nu\circ M$.
\end{enumerate}
Moreover, if $(Y,{\mathcal B})$ is a standard Borel space, then the conditions
are also equivalent to
\begin{enumerate}
\item[{\rm(ii')}] $P(B)=M(f^{-1}(B))$, $\forall B\in {\mathcal B}$,
with $f:X\to Y$ measurable,
\item[{\rm(iii')}] there is a Markov kernel $\lambda$, such that
$P=\lambda\circ M$.
\end{enumerate}
\end{theorem}

We remark that the equivalence (i) $\Leftrightarrow$ (ii') for real observables
was proved in \cite{DLPY},  where the proof was based on the Naimark theorem.

\section{Clean observables}

Let $M:(X,\mathcal A)\to \mathcal E(H)$ and $N:(Y,\mathcal B)\to \mathcal E(H)$
be two observables. We  write $M\preceq N$ if there exists a weak Markov kernel $\nu$
with respect to $M$, such that $N=\nu\circ M$. If also $N\preceq M$,
we write $M\sim N$. This defines an equivalence relation on the set of
observables and $\preceq$ is a partial order on the equivalence classes.
Minimal elements with respect to this order are called {\it clean}.

This equivalence and order have a statistical interpretation: if $M\preceq N$,
 then the family of probability measures $\mathcal P_M:=\{m\circ M\, :\, m\in\mathcal S\}$
 is more informative than $\mathcal P_N:=\{m\circ N\, :\, m\in \mathcal S\}$,
 in the sense  that  the elements of $\mathcal P_M$ can be distinguished more
 precisely by statistical procedures than elements of $\mathcal P_N$, \cite{St}.
 We remark that previous  definitions of $\preceq$ involved smearings with respect to Markov kernels
 rather than  weak Markov kernels.
 In the case of standard Borel spaces, the two notions are equivalent, whereas in the general
 situation, the weaker definition seems to be more appropriate.

The results of the previous sections can be applied to the
characterization of cleanness of sharp  observables. For this, we
need the following simple observation.

\begin{lemma}\label{le:proj} If a projection $P$ is contained in the
range of an observable $M$, then $P$ commutes with all elements of
${\mathcal R}(M)$.
\end{lemma}
\begin{proof}
Let $P=M(A)$ for a set $A$ and let $M(B)\in {\mathcal R}(M)$. From
$M(A\cap B)\leq M(A)$, and since $M(A)$ is a projection, we have
$M(A\cap B)=M(A\cap B)M(A)=M(A)M(A\cap B)$. Similarly, $M(A^c\cap
B)=M(A^c\cap B)M(A^c)=M(A^c)M(A^c\cap B)$. From this
$M(A)M(B)=M(A)M(A\cap B)+M(A)M(A^c\cap B)=M(A\cap B)=M(B)M(A)$.
\end{proof}

\begin{corollary}\label{co:clean} A PV measure is clean iff its range
generates a maximal abelian von Neumann subalgebra of ${\mathcal
B}(H)$.
\end{corollary}
\begin{proof} Let $E$ be a PV measure. Let the abelian von Neumann
subalgebra ${\mathcal N}$ generated by ${\mathcal R}(E)$ be not
maximal. Then there is a  maximal abelian von Neumann subalgebra
${\mathcal M}$ which contains ${\mathcal N}$ and a PV measure $F$,
such that ${\mathcal R}(F)$ generates  ${\mathcal M}$. Then
${\mathcal R}(E)\subset {\mathcal R}(F)$. By Theorem \ref{th:equiv},
$E$ is a smearing of $F$, but $F$ is not a smearing of $E$.
Therefore $E$ is not clean.

Assume that ${\mathcal R}(E)$ generates a maximal abelian von
Neumann subalgebra ${\mathcal M}$. Since $H$ is separable, there is
a self-adjoint operator $T$ such that ${\mathcal R}(T)={\mathcal
R}(E)$, and ${\mathcal M}=\{ T\}''$.  In particular, every
projection $P$ in ${\mathcal M}$ belongs to ${\mathcal
R}(T)={\mathcal R}(E)$.

Suppose that  $E$ is a smearing of a POV measure $M$. Then
${\mathcal R}(E)\subset {\mathcal R}(M)$, and \newline
$E(B)M(C)=M(C)E(B)$ for all $B,C$ by Lemma \ref{le:proj}. Therefore
${\mathcal R} (M)\subset {\mathcal M}'={\mathcal M}=\{T\}''$. As in
the proof of Theorem \ref{thm:commut}, this implies that $M$ is a
smearing of $E$.
\end{proof}

\end{document}